\documentclass[epj]{svjour}
\usepackage{latexsym,amsmath}
\usepackage{epsfig}

\begin{document}

\title{Efficient Stark deceleration of cold polar molecules}

\author{Eric R. Hudson, J.~R. Bochinski, H.~J. Lewandowski, Brian C. Sawyer, \and Jun Ye}
\authorrunning{Eric R. Hudson et. al}
\titlerunning{Efficient Stark deceleration of cold polar molecules}
\mail{Eric.Hudson@colorado.edu}
\institute{JILA, National Institute of Standards and Technology and\\ University of Colorado and Department of Physics, University of Colorado,\\ Boulder, Colorado 80309-0440}
\date{Received: date / Revised version: date}

\abstract{Stark deceleration has been utilized for slowing and trapping several species of neutral, ground-state polar molecules generated in a supersonic beam expansion. Due to the finite physical dimension of the electrode array and practical limitations of the applicable electric fields, only molecules within a specific range of velocities and positions can be efficiently slowed and trapped. These constraints result in a restricted phase space acceptance of the decelerator in directions both transverse and parallel to the molecular beam axis; hence, careful modeling is required for understanding and achieving efficient Stark decelerator operation.  We present work on slowing of the hydroxyl radical (OH) elucidating the physics controlling the evolution of the molecular phase space packets both with experimental results and model calculations.  From these results we deduce experimental conditions necessary for efficient operation of a Stark decelerator.
} 
\maketitle

\section{Introduction}
\label{intro}
The study of ultra-cold polar molecules is a rapidly emerging field as evidenced by this special journal edition, and the large number of new experiments being undertaken \cite{1,2,3,4,5,6}.  The permanent electric dipole moment, $\mu$, possessed by polar molecules allows for a new type of interaction in the ultra-cold environment.  This electric dipole-dipole interaction (and control over it) should give rise to unique physics and chemistry including novel cold collision dynamics \cite{7}, quantum information processing \cite{8}, and a second order phase transition analogous to spontaneous magnetization of a ferromagnet.  When the translational energy of colliding polar molecules becomes comparable to their dipole-dipole interaction energy, the molecules can dramatically influence each other's trajectory.  Lack of spherical symmetry in the interaction causes colliding molecules to be attracted or repelled depending on their relative orientation.  Thus, an external electric field, which orients the molecules, may have a profound effect on the molecular interactions leading to large changes in observed scattering rates \cite{7}.  Another interesting possibility is the observation of criticality in a two or three (3D) dimensional lattice of dipolar molecules.  For a typical polar molecule ($\sim$ 2 Debye) in a 3D optical lattice with 1 m spacing, the critical temperature for the second order phase transition (spontaneous polarization) is estimated by a 3D Ising model as T$_c \sim 3 \mu^2/(2 \pi \epsilon_0  k_B R^3) = $ 200 nK, where \textit{R} is the dipole-dipole separation,  $\epsilon_0$ is the permittivity of free space, and $k_B$ is Boltzmann's constant \cite{9}.  By using extremely polar molecules such as those predicted to occur from photo-association of free radicals with Rubidium ( $\mu \sim$ 10 Debye) \cite{10}, or by decreasing the dipole-dipole separation through simply trapping polar molecules at a sufficient density, the critical temperature may be raised by at least one order of magnitude.  It is interesting to note that at temperatures significantly higher than this regime an analog to adiabatic demagnetization refrigeration \cite{11} may become possible, leading to a new method for further cooling polar molecules, so-called ``paraelectric cooling'' \cite{12}.

The technique of Stark deceleration is especially well-suited for production of cold polar molecules since it utilizes the Stark shift associated with polar molecules to remove the mean translational energy from a supersonic beam of molecules.  Supersonic molecular beams, utilized extensively in physical chemistry, are capable of producing intense pulses of molecules with mean velocities of a few hundred meters per second and a small velocity spread about this mean.  Therefore, in the moving frame of the molecular pulse the associated velocity spread corresponds to a low temperature distribution ($\sim$ 1 K).  Stark deceleration conservatively removes this mean velocity, leaving behind a cold molecule distribution in the laboratory frame.  To date this technique has been employed to produce slow, cold beams of CO \cite{2}, OH \cite{3}, ND3 \cite{13}, resulting in a trapped sample of 10$^4$ molecules at 25 mK in the case of ND$_3$.  Slowing of YbF in an alternate-gradient decelerator has also been demonstrated in a proof of principle experiment \cite{6}.

Due to practical limitations of both the physical dimensions of a Stark decelerator and the applicable electric fields, only molecules within a specific range of velocities and positions can be efficiently slowed and trapped.  These constraints result in a restricted phase space acceptance of the decelerator in directions both transverse and parallel to the molecular beam axis.  Thus, for efficient operation care must be taken to match the phase space distribution of the supersonic beam source (emittance) to the phase space acceptance of the Stark decelerator.  This article details work from both experiment and model, describing the process of phase space matching between the supersonic beam source and the molecular decelerator.  The focus of the current article is to present a clear understanding of the decelerator efficiency and describe explicitly the experimental requirements of maximizing it.  For experimenter new to this field, this article will serve as a useful and practical guide for design of future Stark deceleration experiments.  

Our experiment centers on the deceleration of the neutral hydroxyl radical, OH.  For low rotational levels the $^2\Pi$  electronic ground state of OH is sufficiently described by Hund's case (a).  Spin-orbit coupling results in the $^2\Pi_{3/2}$ state lying $\sim$139 cm$^{-1}$ below the $^2\Pi_{1/2}$ state.  Because of the non-zero orbital angular momentum of the unpaired electron in OH each total angular momentum, J, state is  $\lambda$-type doubled, resulting in two closely spaced ($\sim$1.5 GHz), opposite parity levels in the ro-vibrational ground state.  The application of an electric field readily mixes these states and for the symmetric f state results in an increase in energy with electric field \cite{14}.  The most polarized sub-level of this state, denoted as $|^2\Pi_{3/2}$, J = 3/2, m$_J$ = 3/2, f $\rangle$, is decelerated by our Stark decelerator.  We detect the presence of OH molecules through the technique of laser induced fluorescence.  A pulsed excitation laser tuned to 282 nm promotes the molecules along the A$^2\Sigma_{1/2}(v=1, J=3/2)\leftarrow$X$^2\Pi_{3/2} (v=0, J=3/2)$  electronic transition, excited molecules subsequently decay primarily (73\%) back to the ground electronic state along the A$^2\Sigma_{1/2}
(v = 1)\rightarrow$X$^2\Pi_{3/2} (v = 1)$ pathway at 313 nm.  This fluorescence is collected and imaged onto a photomultiplier tube.  By varying the time of the laser pulse, we sample the OH molecules at a single location at different times to extract time-of-flight (ToF) information.  Careful measurements of the total molecule numbers before and after the Stark decelerator, and thus determination of the overall decelerator efficiency as a function of decelerator/supersonic beam operation parameters, enables information pertaining to the phase space matching of the source to the decelerator to be obtained. We have developed a simple model along with results of Monte Carlo simulations that provide excellent agreement with the observed behavior.  The remainder of the article is organized as follows.  Section 2 briefly describes the operation of a Stark decelerator (for a more complete review of Stark deceleration see \cite{15,16}) as well as develops the simple intuitive model for phase space matching.  Section 3 details our experimental results and compares them to the model of Section 2 to provide confidence in using it for optimization, while Section 4 is reserved for conclusions and suggestions for optimal operation of a Stark decelerator.

\section{Stark Deceleration}
\label{sec:1}
Figure 1 shows a schematic of our Stark decelerator, which has been described elsewhere\cite{16}.  Hydroxyl radical molecules produced by the discharge of H$_2$O in Xenon ($\sim$1:99 mixture) undergo a supersonic expansion and are subsequently skimmed to separate the low vacuum source region ($\sim$10$^{-4}$ torr) from the high vacuum region ($\sim$10$^{-7}$ torr), which contains the decelerator electrodes and the applied high-voltage electric fields.  After skimming, the molecules are focused by an electrostatic hexapole field to provide transverse coupling into the Stark decelerator.  Once the molecules are coupled into the Stark decelerator, the slowing process begins.  The Stark decelerator is constructed of 69 slowing stages spaced 5.475 mm apart with each stage comprised of two cylindrical electrodes of diameter 3.175mm separated axially by a 5.175 mm and oppositely biased at high voltage ($\pm$ 12.5 kV).  Successive stages are oriented orthogonally to each other to provide transverse guiding of the molecular beam.  The geometry of the slowing stages provides an electric field maximum between the electrodes with the field decreasing away from the electrode center as seen in Figure 2(a).  Therefore, a weak-field seeking molecule travelling longitudinally down the decelerator will be decelerated as it moves into the region between the electrodes and will remain decelerated if the high voltage is removed before the molecule passes through the electrode's field.  In order to minimize the number of high voltage switches used, all the like-oriented electrodes of the same polarity are connected to one switch (4 switches total for the entire decelerator).  

\begin{figure}
\leavevmode \epsfxsize=3.375in \epsffile{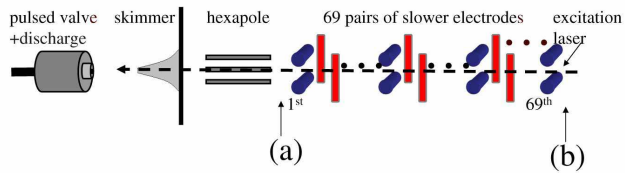}
\caption{\label{fig1}Schematic of the Stark decelerator, displaying the pulsed valve and discharge plates, the molecular beam skimmer, the electric hexapole, and the electrode stages.  The electrode stages alternate orientation (vertical - horizontal) as shown in the figure.  The spatial locations indicated by arrows correspond to the locations where molecule number is measured to determine decelerator efficiency. }
\end{figure}

\begin{figure}
\epsfxsize=3.375in \epsffile{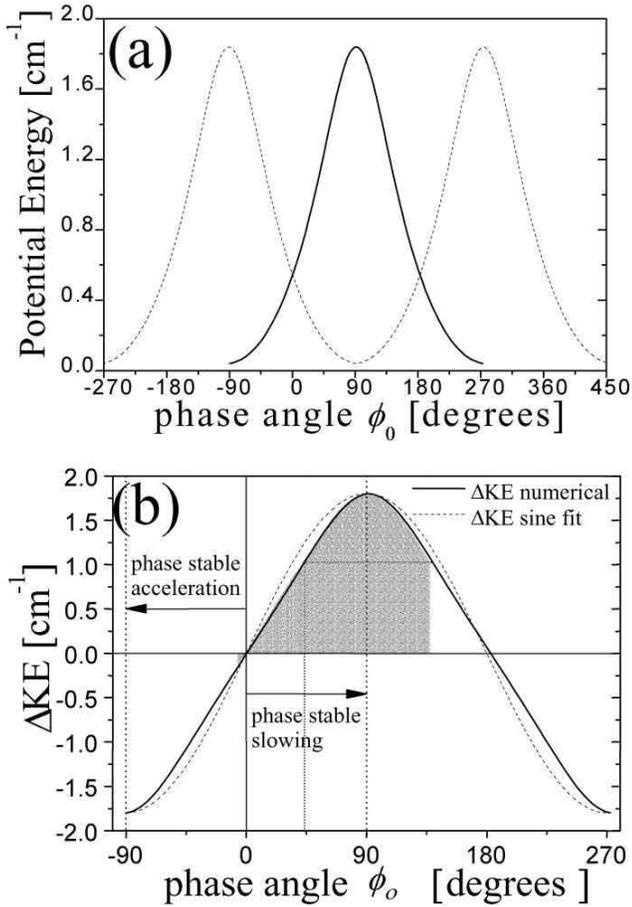}
\caption{\label{fig2}Phase stable operation of the decelerator. (a) Longitudinal Stark energy potentials generated by the two sets of electrodes, where the solid line represents the potential from one set of electrodes and the dotted line from the other set. (b) Kinetic energy loss per stage ($\Delta$KE) experienced by molecules from switching between the two potentials given above. Dashed line corresponds to a sine function approximation of the potential energy loss.  The shaded region corresponds to the phase stable area when decelerating at 45$^\circ$.}
\end{figure}

As a consequence of this switch minimization, when the voltage at each slowing stage is grounded and the next stage is turned on, the electric field is not completely removed, but rather becomes the field generated by the alternate set of electrodes (i.e. the solid versus dashed curves in Figure 2(a)).  Thus, the decrease in the molecular kinetic energy, $\Delta$KE, is given as the difference in potential energy generated by the two sets of electrodes and is shown as a function of synchronous molecule phase angle, $\phi_0$, in Figure 2(b). The synchronous molecule phase angle, $\phi_0$is defined as:
\begin{equation}
\phi_0 = \frac{180^\circ}{L}z,
\label{eqn1}
\end{equation}
where \textit{L} is the distance between two adjacent stages and the zero position of \textit{z} is defined  to be exactly between two adjacent stages.  Thus, switching at $\phi_0$ = 0$^\circ$ corresponds to no net change in molecular kinetic energy, while maximum deceleration (acceleration) occurs at $\phi_0$ = 90$^\circ$ (-90$^\circ$).  Though synchronous molecule phases between $\phi_0$ = 0$^\circ$ and 180$^\circ$ lead to deceleration, only 0$^\circ < \phi_0 < 90^\circ$ results in stable slowing because it is only on the positive sloping $\Delta$KE curve that molecules ahead (behind) of the center of the bunch are decelerated more (less) than the average deceleration.  To further investigate the dynamics of the molecules within this stable packet about the synchronous molecule (\textit{i.e.} the center of the bunch) it is useful to define the excursion of a molecule from the synchronous molecule position as $\Delta\phi = \phi-\phi_0$. Then with the use of Newton's second law and the sine function fit to the change of kinetic energy as function of  $\phi_0$ (Figure 2(b)) it is trivial to write \cite{17,15}:
\begin{equation}
\frac{d^2\Delta\phi}{dt^2}+\frac{\pi C_{Max}}{mL^2}(\sin(\Delta\phi+\phi_0)-\sin(\phi_0))=0,
\label{eqn2}
\end{equation}
where $C_{Max}$ is the maximum kinetic energy change per slowing stage, \textit{t} is the time coordinate, and \textit{m} is the molecular mass.  This equation is that of a harmonic oscillator with its equilibrium position offset (represented by $\phi_0$).  Thus one expects non-synchronous molecules to oscillate around the synchronous molecule position inside an asymmetric oscillator potential.  From numerical integrations of Eq. (2) and the first time derivative of Eq. (1) one can solve for the stable and unstable regions of phase space, shown as a function of the synchronous molecule phase angle in Figure 3 for OH under typical decelerator operation.  The most important feature of Figure 3 is the rapidly decreasing area of stable evolution (i.e. region bounded by the separatrix).  This decrease in stable area is easily understood by analogy to the pendulum driven by a constant torque.  As the torque is increased (\textit{i.e.} the synchronous phase angle is increased) the equilibrium position of the pendulum is pushed toward the apex, therefore the amplitude of oscillations that result in stable oscillatory motion is reduced.  This behavior is responsible for the separatrix possessing only one cusp for non-zero  $\phi_0$.  It is clear that the number of molecules accepted by a decelerator operating at a specific synchronous molecule phase angle is then given as an area integral inside the separatrix weighted by the supersonic beam distribution at the decelerator's entrance.

This phase space ``bucket'' loading is illustrated graphically in Figure 4.  In the first panel of this figure, ToF data taken at the decelerator entrance (location ``a'' indicated in Figure 1) are shown along with a hypothetical ideal supersonic beam of OH molecules (v$_{z,center}$ = 300 m/s,  v$_z$ = 30 m/s, where $v_{z,center}$ is the pulse mean speed and $\Delta v_z$ is the full width at half maximum of the distribution (FWHM)).  For this graph the supersonic molecular beam parameters were varied by tuning the discharge initiation time as described previously \cite{18} resulting in the ability to input two vastly different molecular beams into the decelerator.  The first distribution utilized has v$_{z,center}$  = 415 m/s with $\Delta v_z$ of $\sim$ 90 m/s, while the second distribution is centered about 350 m/s with a spread of $\sim$ 80 m/s.  For these two experimental distributions and the idealized case, Monte Carlo simulation results are used to construct graphs of the longitudinal phase space occupied by the molecules in the remaining panels of the figure.  Each gray point in these graphs represents the location in phase space of a sample molecule as predicted by our simulations at the time of loading into the decelerator.  The two dashed curves on each graph are projections of the phase space distribution onto the respective axes.  Overlaid on these graphs are the separatrix curves for operation at $\phi_0$ =0$^\circ$, the so-called bunching condition, and for slowing the molecules to rest at the exit of the decelerator.  From these graphs, the importance of creating a cold (\textit{i.e.} small $\Delta$v) source of molecules as an input for a Stark decelerator is clear; with a smaller $\Delta$v the molecular pulse spreads less in route from generation at the source to the decelerator entrance, and thus is more efficiently coupled into the decelerator in both the spatial and velocity dimensions.  The beneficial effect of lowering the pulse mean speed is also made evident; as the mean speed drops, the required phase angle for slowing is reduced and thus the separatrix area increases. 

\begin{figure}
\epsfxsize=3.375in \epsffile{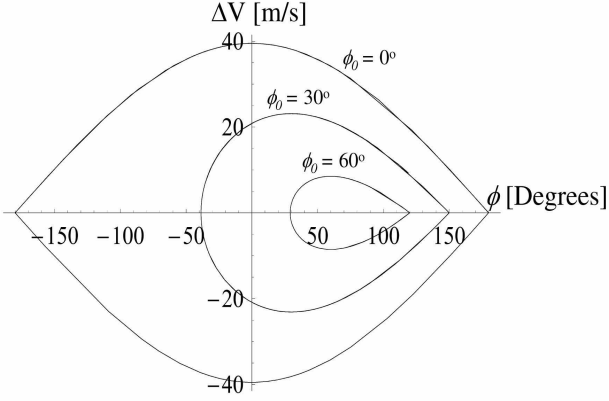}
\caption{\label{fig3}Phase stable area (separatrix area) versus phase angle for $\phi_0$ = 0$^\circ$, 30$^\circ$, and 60$^\circ$.  As the phase angle is increased, the stable area is reduced; equivalently, the stable longitudinal velocity width narrows, and assuming the source completely populates the phase stable region, decreases the number of molecules in the stable packet. }
\end{figure} 

\begin{figure}
\epsfxsize=3.375in \epsffile{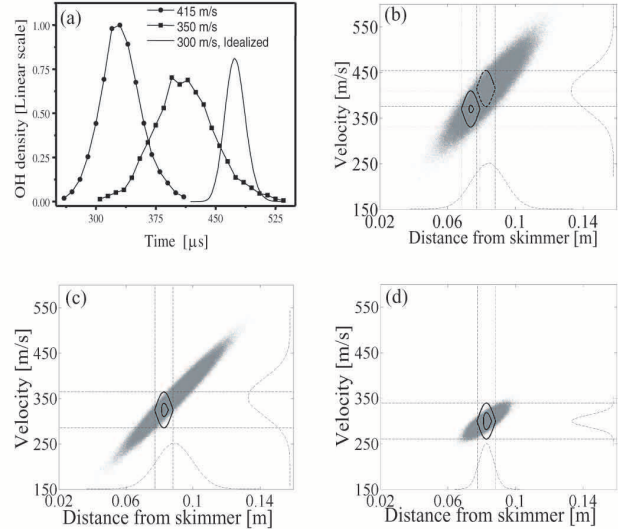}
\caption{\label{fig4} Longitudinal bucket loading.  Panel (a) represents time-of-flight measurements taken at the decelerator input (position ``a'' in figure 1) for different molecular beam operating conditions.  Filled circles (filled squares) represent operating with a distribution with a mean speed of 415 m/s (350 m/s) and a full width at half maximum spread of 95 m/s (85 m/s), while the solid line is an idealized molecular beam of 300 m/s mean speed with 10\% velocity spread.  Panels (b), (c), and (d) represent the longitudinal phase space at the beginning of the Stark deceleration as derived from our Monte Carlo simulations.  In panel (b) the dashed separatrix represents bunching for 415 m/s, while the larger solid line is the separatrix for bunching at 370 m/s.  Inside the 370 m/s bunching separatrix is shown the separatrix for slowing a packet of molecules to 25 m/s.  In panel (c) the separatrix are shown for bunching at 325 m/s and slowing to 25 m/s molecules from the 350 m/s distribution, while panel (d) shows bunching at 300 m/s and slowing to 25 m/s for the hypothetical molecular beam.}
\end{figure}

As aforementioned, the effects of supersonic beam and Stark decelerator parameters on overall efficiency can be estimated from knowledge of the separatrix overlap with the supersonic beam phase space.  Approximating the area enclosed by the separatrix as that of the corresponding rectangle whose height (width) correspond to the separatrix velocity (spatial) bounds and ignoring the minimal convolution due to the short free flight of the molecules after creation, the overall longitudinal slower efficiency is given as:
\begin{gather}
\eta_{Longitudinal} = N[\nu_z, \Delta \nu_z, \nu_{z, Bounds}(\phi_0)]\nonumber \\\times N[z, \Delta z, z_{Bounds}(\phi_0)],
\end{gather}
with

\begin{equation}
N(x, \Delta x, x_{Bounds}) = \frac{\int_{X Lower bound(\phi_0)}^{X Upper bound(\phi_0)}f(x, \Delta x) dx}{\int_{-\infty}^\infty f(x, \Delta x)dx},
\label{eqn3b}
\end{equation}

\noindent where \textit{z} refers to the dimension along the decelerator's axis (longitudinal axis) and $f(x, \Delta x$) is the molecular distribution in the appropriate dimension.  The integration bounds correspond to the separatrix maxima along their appropriate dimension and are an explicit function of the synchronous molecule phase angle.  From Figure 2(b) we see that for $\phi_0$ = 45$^\circ$, non-synchronous molecules with phase angles between 45$^\circ$ and 135$^\circ$ will be slowed more than the synchronous molecule (shaded region).  In general, we see the maximum stable forward excursion is then given as: 

\begin{equation}
\Delta\phi_{Max,+}(\phi_0) = 180^\circ - 2 \phi_0
\label{eqn4}
\end{equation}

Unfortunately, a closed form solution for the maximum backwards excursion, $\Delta \phi_{Max,-}(\phi_0)$, does not exist, however, it can easily be found numerically or estimated by the negative of Eq. (5) since both the maximum stable forward and backward excursions have the same average derivative as a function of  $\phi_0$.  From Eq. (2) we find the work done in bringing a molecule starting at $\Delta \phi_{Max,+}(\phi_0)$  with the same speed as the synchronous molecule (\textit{i.e.}at the separatrix cusp) back to the synchronous molecule position (\textit{i.e. }  $\Delta\phi$ = 0), W, as:
 
\begin{equation}
W = \frac{-C_{Max}}{\pi}\int_{\Delta\phi_{Max,+}(\phi_0)}^0 (\sin(\Delta\phi+\phi_0)-\sin(\phi_0))d\Delta\phi
\label{eqn5}
\end{equation}

Clearly, this molecule will possess the maximum stable velocity and we thus have the separatrix velocity bound as:

\begin{equation}
\Delta\nu_{Max}(\phi_0)= 2\sqrt{\frac{C_{Max}}{m\pi}(\cos(\phi_0)-(\frac{\pi}{2}-\phi_0)\sin(\phi_0))}.
\label{eqn6}
\end{equation}

Now, assuming the molecular distributions are adequately described by the form:
    
\begin{equation}
f(x, \Delta x) = exp\left[ \left(\frac{x-x_{center}}{\frac{\Delta x}{2\sqrt{ln(2)}}}\right)^2\right]
\label{eqn7}
\end{equation}
    
where  $\Delta x$ refers to FWHM and x$_{center}$ to the value at which the distribution is centered we see Eq. (3) becomes:

\begin{gather}
\eta_{Longitudinal} = \nonumber \\  \frac{1}{4}[Erf\left(\frac{(v_{z,design}-v_{z,center})+\Delta\nu_{z, Max}(\phi_0)}{\frac{\Delta\nu_z}{2}}\sqrt{ln(2)}\right) \\ \nonumber -Erf\left(\frac{(v_{z,design}-v_{z,center})-\Delta\nu_{z, Max}(\phi_0)}{\frac{\Delta\nu_z}{2}}\sqrt{ln(2)}\right) ] \\ \nonumber  \times [ Erf\left(\frac{(z_{design}-z_{center})+
\frac{L}{\pi}\Delta\phi_{Max,  +}(\phi_0)}{\frac{\Delta z}{2}}\sqrt{ln(2)}\right) \\ -Erf\left(\frac{(z_{design}-z_{center})-
\frac{L}{\pi}\Delta\phi_{Max, +}(\phi_0)}{\frac{\Delta z}{2}}\sqrt{ln(2)}\right)] \nonumber 
\end{gather}

Where \textit{Erf }refers to the Error function, and v$_{z,design}$ and z$_{design}$ have been introduced to account for the possibility of designing the decelerator pulse sequence to select molecules which are not at the peak of the molecular distribution.  If molecules are loaded at the peak of the distribution (\textit{i.e.} v$_{z,design}$ = v$_{z,center}$ and z$_{design}$ = z$_{center}$) and we approximate  $\Delta\phi_{Max,-}(\phi_0)   = -\Delta\phi_{Max,+}(\phi_0)$, we recover the simple result of our earlier work \cite{15}:

\begin{gather}
\eta_{Longitudninal}= Erf\left(\frac{\Delta\nu_{z,Max}(\phi_0)}{\frac{\Delta\nu_s}{2}}\sqrt{ln(2)}\right) \\
 \times Erf\left(\frac{L(1-\frac{\phi_0}{\pi/2})}{\frac{\Delta x}{2}}\sqrt{ln(2)}\right),\nonumber
\end{gather}

which is useful for estimating expected decelerator efficiency.  Plots of Eq. (8) with loading at the peaks of the distributions are shown in Figure 5 versus both $\Delta v_z$ (Figure 5(a)) and  v$_{z,center}$ (Figure 5(b)), where the value of  $\Delta\phi_{Max,-}(\phi_0)$ has been found numerically.  In these graphs the slowing phase angle was chosen to bring the molecules to rest at the decelerator's exit.  Again the importance of a low central velocity and small spread about this mean is evident.  The most striking feature of these graphs is the flattening of the longitudinal slower efficiency for small velocity spreads.  In fact for a sufficiently narrow velocity spread the mean speed of the pulse becomes to some degree ``unimportant'' as the supersonic beam emittance fits entirely inside the decelerator acceptance for most operating conditions.  It is important to note that similar effects can be achieved by changing the characteristics of the Stark decelerator.  By increasing the stage-to-stage spacing one enlarges the spatial bounds of the separatrix resulting in more decelerated molecules; however, increasing the distance without increasing the applied electric field will result in worse transverse guiding and a smaller stable phase space area transverse to the molecular beam axis.  Increasing only the applied electric field results in the ability to decelerate molecules with larger velocity spreads relative to the synchronous molecule speed, and thus increases the overall slower efficiency.  However, the technical challenges associated with going to voltages higher than the present operating conditions are large. Therefore a more practical alternative maybe to increase the number of slowing stages to reduce the required phase angle.

\begin{figure}
\epsfxsize=3.375in \epsffile{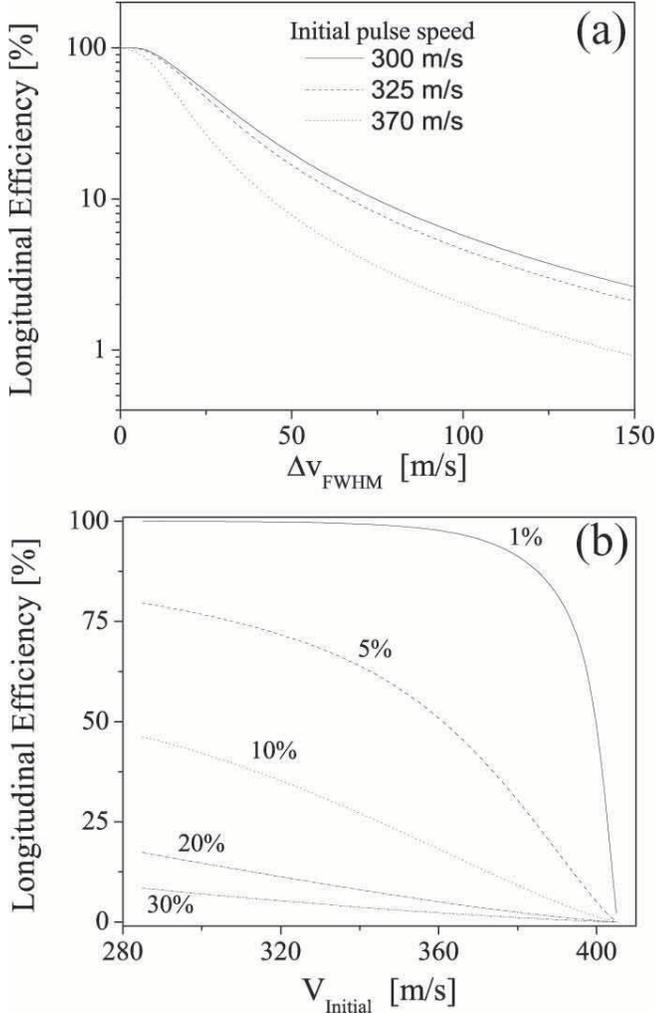}
\caption{\label{fig5} Simple model for longitudinal bucket loading.  Panel (a) shows the effect of varying the supersonic beam's velocity spread on longitudinal efficiency for several initial speeds, while panel (b) shows the longitudinal efficiency as a function of initial pulse speed for several velocity spreads.}
\end{figure}

    Efficient operation of a given decelerator/supersonic beam combination occurs when Eq. (8) is maximized at the desired final speed.  Since the molecular distribution and the desired final speed are assumed fixed the only parameter which can be varied is v$_{z,design}$, the initial speed for which the decelerator pulse sequence is designed.  By decreasing v$_{z,design}$ the required slowing phase angle is decreased and the separatrix area increased. However, if v$_{z,design}$ is lowered too far below v$_{z,center}$ the molecular beam phase space overlaps poorly with the decelerator input separatrix and slower efficiency suffers.  Thus, the most efficient slower operation will occur for a v$_{z,design}$ below v$_{z,center}$ by an amount that depends on the $\Delta v_z$ of the pulse and the desired final speed (typically 1 to 10 percent of v$_{z,center}$ for slowing to rest).  Figure 6 shows the results of maximizing Eq. (8) for the given experimental distributions as well as the idealized source.  In this graph the amount of velocity detuning from v$_{z,center}$ for most efficient operation is shown as a function of desired final speed.  For final packet speeds close to the initial pulse speed (right part of graph) it is most efficient to utilize a phase angle of zero (bunching) because of its large acceptance and vary v$_{z,design}$.  However, as the final velocity is decreased, this procedure eventually leads to a mismatch between the supersonic beam emittance and the decelerator acceptance, and it becomes more efficient to decelerate molecules nearer the center of the supersonic beam distribution, leading to the sharp ``transition'' peaks.  The position of this transition for our decelerator is at a detuning equal to approximately 15\% of the pulse spread and is evident in the figure by the sudden drop in velocity detuning.  As slower molecules are desired, the effect of the decreasing acceptance of the decelerator (with increasing  $\phi_0$) is combated by selecting slower initial molecules, which require less slowing.  This effect is evident by the increase of detuning with slower final speeds. The gain in efficiency by optimizing decelerator operation according to Eq. (8) depends sensitively on both the molecular beam and decelerator parameters, but in the case of our experimental conditions ranges from only a few percent to well over a factor of 100 improvement \cite{19}.

    While Eq. (8) describes the efficiency of decelerator operation in the longitudinal dimension, it says nothing about evolution transverse to the decelerator axis.  In this dimension the molecular distribution is centered about zero in both space and velocity thus Eq. (3) becomes:

\begin{gather}
\eta_{Transverse}=N[0,\Delta\nu_x, \nu_{x,Bounds}(\phi_0)]N[0, \Delta x, x_{Bounds}(\phi_0)]\nonumber \\ N[0,\Delta\nu_y, \nu_{y,Bounds}(\phi_0)]N[0, \Delta y, y_{Bounds}(\phi_0)],
\end{gather}

where x and y refer to the dimensions perpendicular to the decelerator's axis.  Assuming distributions of the form of Eq. (7) we have:

\begin{gather}
\eta_{Transverse}= Erf\left(\frac{\Delta\nu_{x,Max}(\phi_0)}{\frac{\Delta\nu_x}{2}}\sqrt{ln(2)}\right)\nonumber\\ \times Erf\left(\frac{\Delta x(\phi_0)}
{\frac{\Delta x}{2}}\sqrt{ln(2)}\right)Erf\left(\frac{\Delta\nu_{y,Max}(\phi_0)}{\frac{\Delta\nu_y}{2}}\sqrt{ln(2)}\right)\nonumber\\ \times Erf\left(\frac{\Delta y(\phi_0)}
{\frac{\Delta y}{2}}\sqrt{ln(2)}\right)
\end{gather}

\begin{figure}
\epsfxsize=3.375in \epsffile{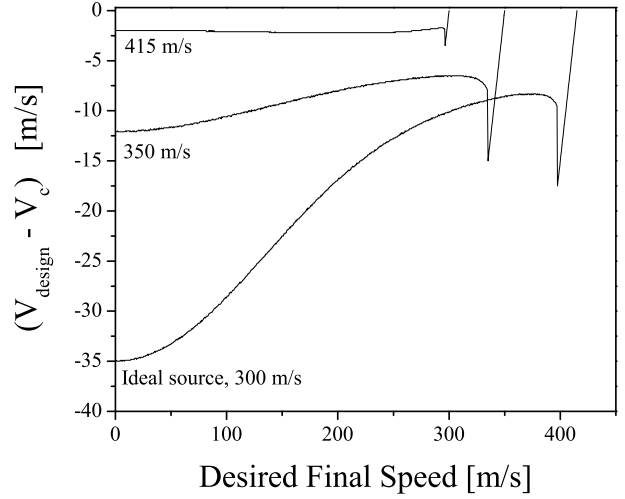}
\caption{\label{fig6}Decelerator optimization condition.  The optimum detuning of the design speed from the pulse's mean speed as a function of desired final speed shown for three different supersonic beam distributions.  }
\end{figure}

The spatial acceptance of the decelerator is set by the decelerator rod spacing, thus $\Delta$x and  $\Delta$y $\sim
$ 1 mm.  Because the transverse evolution of molecules through the Stark decelerator is a complicated dynamical process that depends sensitively on the operating conditions it is not possible to give a simple expression for  $\Delta v_{x,Max(\phi_0)}$ and  $\Delta v_{x,Max( \phi_0)}$, however, for most deceleration experiments to date it can be estimated as $\sim$3 m/s for most slower conditions with minimal error \cite{16,20}.  The behavior of Eq. (11) is analogous to that of Eq. (8), as the velocity (spatial) spread of the source is lowered or the maximum stable velocity (position) is increased by a decelerator change, the efficiency of the slower grows.  It is the aim of proper hexapole focusing to "mode-match" the supersonic beam into this transverse acceptance set by the decelerator.

\begin{figure}
\epsfxsize=3.375in \epsffile{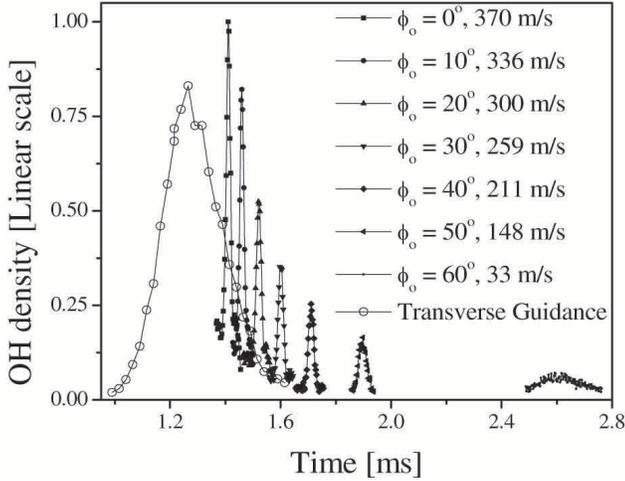}
\caption{\label{fig7}Molecular packets at the decelerator exit (position ``b'' in Figure 1) for varying phase angles.  Peaks arriving later in time are the result of slowed molecules.  From comparison of the number of molecules in these packets to the total inputted molecular number, decelerator efficiency can be determined.}
\end{figure}

\section{Experiment}
Time of flight measurements of OH molecules taken at the exit of the decelerator (position ``b'' in Figure 1) are shown in Figure 7 for various decelerator operating conditions.  Open circles represent operation in transverse guidance mode, where the decelerator array is biased at high voltage, but un-switched resulting in a potential minimum along the decelerator axis.  This mode of operation does not affect the molecules longitudinal speed, but provides transverse guidance through the length of the decelerator.  The remaining traces show the stable packet of molecules produced when switching the decelerator with non-zero phase angle.  As the slowing angle is increased the packets arrive later in time signifying their smaller mean speed.  The spreading of the ToF pulse due to free flight and convolution with the detection window is also evident.  Integration of the deconvolved ToF pulse reveals the total molecular number in each phase stable packet, and is shown in Figure 8, where the data points have been normalized to the bunching condition.  In this graph the total molecular number is plotted versus the final speed of the molecular bunch for the three different operating conditions represented in Figure 4.  The filled circles represent data taken for slowing sequences designed for an initial speed of 415 m/s.  For this data the supersonic beam was operated to yield a distribution centered at 415 m/s with a 90 m/s FWHM.  The filled squares in Figure 8 represent loading molecules from the same distribution into a phase bucket designed for 370 m/s initial speed.  For the data represented by the filled triangles, the supersonic beam was operated to yield a pulse centered at 350 m/s with an 80 m/s spread.  In this data, the decelerator was operated for a 325 m/s initial speed.  The solid lines in this graph represent the simple theory of Section 2, while the open data points correspond to the results of our Monte Carlo simulations.  Despite the assumptions made in deriving Eq. (8) we see the simple theory of Section 2 is quite accurate.  The only noticeable discrepancy occurs for the 415 m/s curve, where it appears that deviation in the experimental bunching point (which is used to normalize the graph) is responsible for the offset.  From this graph we see that operation at 325 m/s is clearly more efficient at producing slow molecules.  Again, the importance of developing a slow cold source as the input for efficient Stark deceleration is clear.  

\begin{figure}
\epsfxsize=3.375in \epsffile{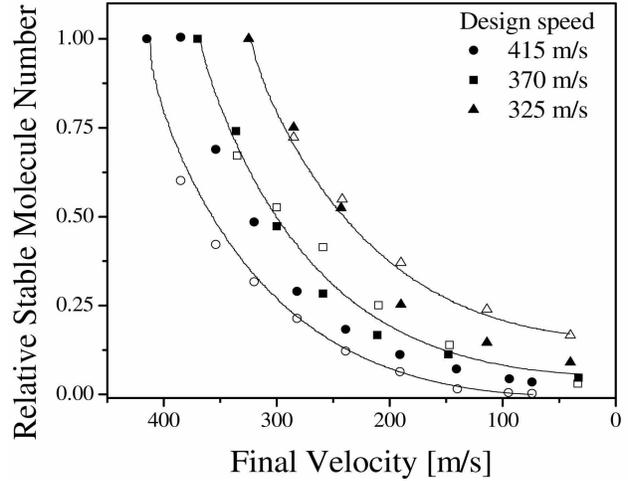}
\caption{\label{fig8} The stable molecule number normalized to $\phi_0$= 0$^\circ$ for three cases of operation.  Circles represent operation with a supersonic beam centered at 415 m/s with a 90 m/s FWHM and the deceleration sequence designed for molecules at 415 m/s.  Squares represent a deceleration sequence designed for 370 m/s molecules from the same beam distribution.  Triangles are for deceleration designed for 325 m/s molecules from a distribution centered at 350 m/s with an 80 m/s spread.  Filled points represent data, while open points represent results from the Monte Carlo simulation.  The solid lines are the results of the simple model of Section 2.}
\end{figure}

By comparing the total molecule number observed in each packet to the total number of molecules inputted into the decelerator (Figure 4 (a)) the overall efficiency of the Stark decelerator can be determined and is shown in Figure 9 for the 370 m/s case as filled squares.  Overlaid on this graph is the simple model of Section 2.  For this theory the transverse parameters  $\Delta$x and  $\Delta$y = 1.5 mm and  $\Delta$v$_x$ and  $\Delta$v$_y$  = 12 m/s   were used in Eq. (11) \cite{21}.  It is remarkable to note that this simple theory predicts the overall slowing efficiency quite accurately with no free parameters.  Clearly, this level of agreement with our simple model gives confidence in using it as an optimization tool.

\begin{figure}
\epsfxsize=3.375in \epsffile{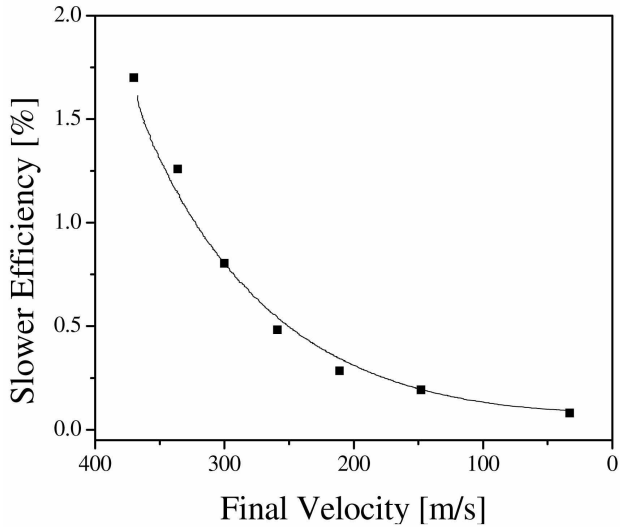}
\caption{\label{fig9}Stark deceleration efficiency.  Efficiency of the Stark decelerator as a function of final speed for slowing designed for an initial speed of 370 m/s taken from the distribution centered at 415 m/s with a 90 m/s FWHM.  Filled squares are data points obtained from the results of Figure 8 normalized by the results of trace (a) in Figure 4.  The solid line is the model of Section 2 including both longitudinal and transverse efficiencies.}  
\end{figure}

\section{Conclusions}

We have developed a simple model of the process of phase space matching between the supersonic beam emittance and the decelerator acceptance.  The ability of this model to predict with accuracy the efficiency of the Stark deceleration process gives confidence in its use as an optimization tool.  While optimization according to this model gives small gain for supersonic sources with low mean kinetic energy and small velocity spreads, it can lead to dramatic improvement for molecular beams that have either a high kinetic energy, a large velocity spread, or both.  As the technique of Stark deceleration is extended to more exotic molecular species, which because of their large mass, internal modes or process of creation result in non-ideal supersonic beams for input into a Stark decelerator, maximization of the decelerator operation efficiency will be essential.

Clearly, it is the aim of the experimenter to maximize the product of Eqs. (8) and (11) for the most efficient decelerator operation.  The efficiency of the decelerator depends sensitively on both the supersonic beam and Stark decelerator parameters.  The importance of developing a cold source as the input for any Stark deceleration experiment cannot be overstated since this is the only real cooling in a slowing experiment and, for a given decelerator, sets the attainable efficiency.  While increasing the dimensions of slowing stages and their separations to allow deceleration of more molecules is a viable way to increase the deceleration efficiency, it is met with increasing technical challenge as the electrode voltage must also be scaled.  Perhaps a more reasonable alternative is to increase the number of slowing stages, thus dropping the amount of energy required to be removed at each stage, resulting in a larger phase space acceptance.  A reasonable design goal is to include enough stages in the decelerator so that molecules can be decelerated to the desired final speed with a phase angle of $\phi_0$ $\leq$ 45$^\circ$.  Also, positioning the source as close as possible to the input of the decelerator is of importance to provide efficient spatial coupling in the longitudinal dimension.  Finally, for a given molecular beam/decelerator combination the optimum phase angle and initial pulse design speed, v$_{z,center}$ , for slowing can be found from Eq. (8).

    This research work is supported by the Keck Foundation, NSF, and NIST. H. J. L. acknowledges support from the N.R.C.


\begin{thebibliography}{}
\bibitem{19}
Since 415 m/s molecules cannot be slowed to rest in our decelerator, detuning the pulse design speed to decelerate molecules with 405 m/s (the maximum speed which can be brought to rest in our machine) or less initial speed would by definition be an infinite gain in efficiency.  Thus the usefulness of this technique for hot sources is evident.
\bibitem{21}
The parameters used for the beam's transverse characteristics are for the gaussian of best fit to the transverse beam profile generated from the Monte Carlo simulations.
\end{thebibliography}
\end{document}